\documentclass[journal=jacsat,manuscript=article, layout=traditional]{achemso}

\usepackage[version=3]{mhchem} 
\usepackage[LGRgreek]{mathastext}
\usepackage[percent]{overpic}
\usepackage{upgreek}
\usepackage{amsmath}

\author{Manasa Kaniselvan}
\email{mkaniselvan@iis.ee.ethz.ch}
\affiliation[ETH Zürich]
{Integrated Systems Laboratory, Department of Information Technology and Electrical Engineering, ETH Zürich, Zürich}

\author{Mathieu Luisier}
\email{mluisier@iis.ee.ethz.ch}
\affiliation[ETH Zürich]
{Integrated Systems Laboratory, Department of Information Technology and Electrical Engineering, ETH Zürich, Zürich}

\author{Marko Mladenovi\'c}
\email{mmladenovic@iis.ee.ethz.ch}
\affiliation[ETH Zürich]
{Integrated Systems Laboratory, Department of Information Technology and Electrical Engineering, ETH Zürich, Zürich}

\title[An \textsf{achemso} demo]
  {An Atomistic Model of Field-Induced Resistive Switching in Valence Change Memory}
  
\begin{document}


\begin{abstract}

In Valence Change Memory (VCM) cells, the conductance of an insulating switching layer is reversibly modulated by creating and redistributing point defects under an external field. Accurate simulations of the switching dynamics of these devices can be difficult due to their typically disordered atomic structures and inhomogeneous arrangements of defects. To address this, we introduce an atomistic framework for modelling VCM cells. It combines a stochastic Kinetic Monte Carlo approach for atomic rearrangement with a quantum transport scheme, both parameterized at the \textit{ab initio} level by using inputs from Density Functional Theory (DFT). Each of these steps operates directly on the underlying atomic structure. The model thus directly relates the energy landscape and electronic structure of the device to its switching characteristics. We apply this model to simulate non-volatile switching between high- and low-resistance states in an TiN/HfO$_2$/Ti/TiN stack, and analyze both the kinetics and stochasticity of the conductance transitions. We also resolve the atomic nature of current flow resulting from the valence change mechanism, finding that conductive paths are formed between the undercoordinated Hf atoms neighboring oxygen vacancies. The model developed here can be applied to different material systems to evaluate their resistive switching potential, both for use as conventional memory cells and as neuromorphic computing primitives.

\end{abstract}


\newpage

\section{Introduction}

Modern computer architectures are increasingly limited by the latency, packing density, and power consumption of their memory. The need for more advanced computing has led to the emergence of a variety of alternative memory cells, including conductive bridging random access memory (CBRAM), phase-change memory (PCM), and valence change memory (VCM). All of these device principles operate by modulating the resistance of a switching layer, typically an oxide, embedded between two electrodes. Collectively, these `resistive switching' devices, also referred to as `memristors', are being explored both to overcome the shortfalls of conventional semiconductor-based non-volatile memory units \cite{AnChen, Baek, Tsunoda2007, Li2015} and as computing primitives for emerging in-memory or neuromorphic computing architectures \cite{Ielmini2018, Hong2018}. Amongst them, VCM cells have the advantage of scalability, ease of fabrication, and large dynamic ranges \cite{Islam2019, Ielmini2018}. 

The operation of a VCM cell can be separated into three distinct states: electroforming, RESET, and SET. In the forming phase, a high bias is applied across the VCM device, inducing a soft dielectric breakdown of the switching layer which significantly increases its conductance. Applying a lower bias of the opposite polarity can recover a lower conductance. The device can thus be cycled between high- (HRS) and low- (LRS) resistance states in RESET and SET processes, within some level of variability in the conductance. On the atomic scale, transitions between each of these states can be explained by the movement of mobile oxygen ions and vacancies through the lattice; the resulting change in the valence of the cations left behind modulates the local conductivity. This mechanism is supported by experimental imaging studies which identify local deformations and bubbling indicating the presence of O$_2$ \cite{JoshuaYang2009, Szot2006}, localized conductive plugs at the electrode interface \cite{Kumar2016, Hubbard2021, Park2013, Li2017-wf}, evidence of reduced oxygen concentrations in the bulk \cite{Hubbard2021, Privitera2013-vc}, and a filamentary increase in conductance through the oxide \cite{Celano2015-nv, Hubbard2021, Privitera2013-vc}. 

However, imaging studies are unable to resolve the dynamics of filament formation or the microscopic effects of process parameters that cannot be easily decoupled. The materials used to build the VCM stack also have a large influence on the HRS-to-LRS ratio and device variability, which are the factors limiting their performance in neuromorphic computing applications \cite{Wang2019, Prezioso2015}. The resistance of each state of a VCM device is also highly dependant on the precise atomic arrangement in the switching layer, which uniquely correlates to a specific distribution of defect energies that can conduct electronic current through the switching layer. All of this indicates a strong dependence of a VCM's performance on the specifics of its atomic and electronic structure. Accurate physical modelling of these effects would aid in optimizing VCM architectures and material compositions. 

Much of the existing work in the simulation of VCM cells has focused on a single scale of their operation. Atomistic-level studies on dielectric breakdown mechanisms in several oxides considered for VCM applications have explored the process of oxygen vacancy formation and diffusion \cite{Gao2019, Schie2017, Stewart2019, Clima2016, OHara2014, Zhao2015} and calculated the accompanying changes in electronic structure. They identify a dependence of activation energies on the local defect environment \cite{Zhao2015}, electric field, grain boundaries, and interface environment \cite{Clima2016, OHara2014}, as well as the energy signature of different defects \cite{Foster2002}. Simulations at the structural level explore possible switching dynamics using Molecular Dynamics (MD) \cite{Aldana2020, Urquiza2021, Xu2020} or Kinetic Monte Carlo (KMC) methods \cite{Aldana2020-pp, Loy2020, Long2013-be}. For many oxide stacks, they generally recover the anticipated switching mechanism in which ion exchanges at an interface coupled with the generation of a percolating path of oxygen vacancies leads to an increase in electrical conductance. Finally, transport simulations have revealed a high sensitivity of the current to the thickness of the conductive filament, down to a single line of atoms \cite{Cartoix2012}. Further analysis of the conduction mechanism in these structures has found two distinct modes of transport: \cite{Funck2021, Funck2018} Schottky barrier modulation and trap-assisted tunneling through defect states. The dominance of each mechanism depends on the distribution of defect-induced energy levels within the bandgap of the oxide. 

Modelling the full switching behavior of these devices at multiple scales is, however, complicated by the presence of both atomic and electronic currents and by their stochastic nature. Existing multiscale models can reproduce the general trends observed in resistive switching \cite{Kim2014, Kim2013, Dirkmann2018, Padovani2015, Padovani2017, Abbaspour2018, Bersuker2011, Kopperberg2021, Zeumault2021, Sadi2016-eb}, but face two limitations. First, they treat current with analytical trap-assisted tunneling models. The corresponding equations are typically derived for single trap energies. Changes in the distribution of defects across the VCM and the existence of defect subbands are difficult to capture with this approach. Second, they use a discretized grid of possible atomic sites rather than a true atomic lattice, and they treat activation energies as phenomenological fitting parameters to experimental results, removing the influence of the underlying atomic structure. This precludes a predictive exploration of the full design space of VCM materials.

As existing device models cannot capture all relevant effects and atomic-scale techniques are unable to extrapolate their influence on device-level characteristics, simulating these devices at all scales of operation requires a different approach. Here we present such a multiscale simulation framework dedicated to VCM. Our method combines two approaches to study the behavior of oxygen vacancy filaments in a VCM cell and the electrical current that flows through them. The first is a stochastic KMC model that uses an atomic structure and energy landscape parameterized by Density Functional Theory (DFT). The final atomic structures from the KMC method are passed to an \textit{ab initio} quantum transport simulation, performed on the same atomic grid. This framework allows for the investigation of how the electronic structure and energy landscape of different materials affect their transport properties and switching performance. As such, it can be used to probe the influence of the electronic structure and energy landscape of different switching materials on their transport properties.

\section{Results \& Discussion}

\subsection{Modelling Framework}

Several oxides have been proposed for VCM applications, including HfO$_2$, TaO$_x$, and AlO$_x$ \cite{Li2017-wf, Park2013, Panda2021}. Amorphous HfO$_2$ is the most widely studied amongst them due to its existing experimental maturity and CMOS compatibility \cite{Banerjee2022}. In several studies, it has been sandwiched between a TiN inactive electrode and Ti active electrode, with additional TiN deposited as an inert capping layer \cite{Walczyk2011, Sowinska2012}. The Ti layer also acts as a trapping layer or `reservoir` to capture the generated ions. To relate to such studies, we chose to investigate a TiN/HfO$_2$/Ti/TiN device in this work.

Current flows in HfO$_2$-based VCM result from trap-assisted tunneling through defect sites created by oxygen vacancies; these defects are assumed to originate from the change in valence of the under-coordination of Hf atoms in the vicinity of these vacancies, leaving them conductive. This mechanism is supported by experimental data revealing hopping-type conduction \cite{Loy2020}, a linear temperature-dependence in the low-resistance state \cite{Privitera2013-vc}, as well as theoretical studies on the location and influence of oxygen vacancies on the density of states (DOS) of HfO$_2$ \cite{Kopperberg2021, Funck2021, Strand2022}. The modelling approach we develop here assumes a distribution of charge in a way that is consistent with the filamentary nature of HfO$_2$ VCM cells.

The framework proceeds in three distinct phases: (1) generating the TiN/HfO$_2$/Ti/TiN atomic structure and energy landscape, (2) modelling the atomic rearrangement under an applied field, and (3) calculating the flow of current through the structure after field-induced changes. It uses several computational tools that implement the different physical processes behind atomic and electronic motions, including MD and DFT, as well as an in-house KMC code and Quantum Transmitting Boundary Method (QTBM) solver. \textbf{Fig.~1} presents a schematic of this model. The steps above are linked through exchanges of quantities such as the device's atomic structure information, potential, and Hamiltonian and Overlap matrices. We present an overview of the model below; further details about each of the calculations can be found in the Methods section.

\begin{figure}[H]
         \centering
         \begin{overpic}[width=\textwidth, tics=10]{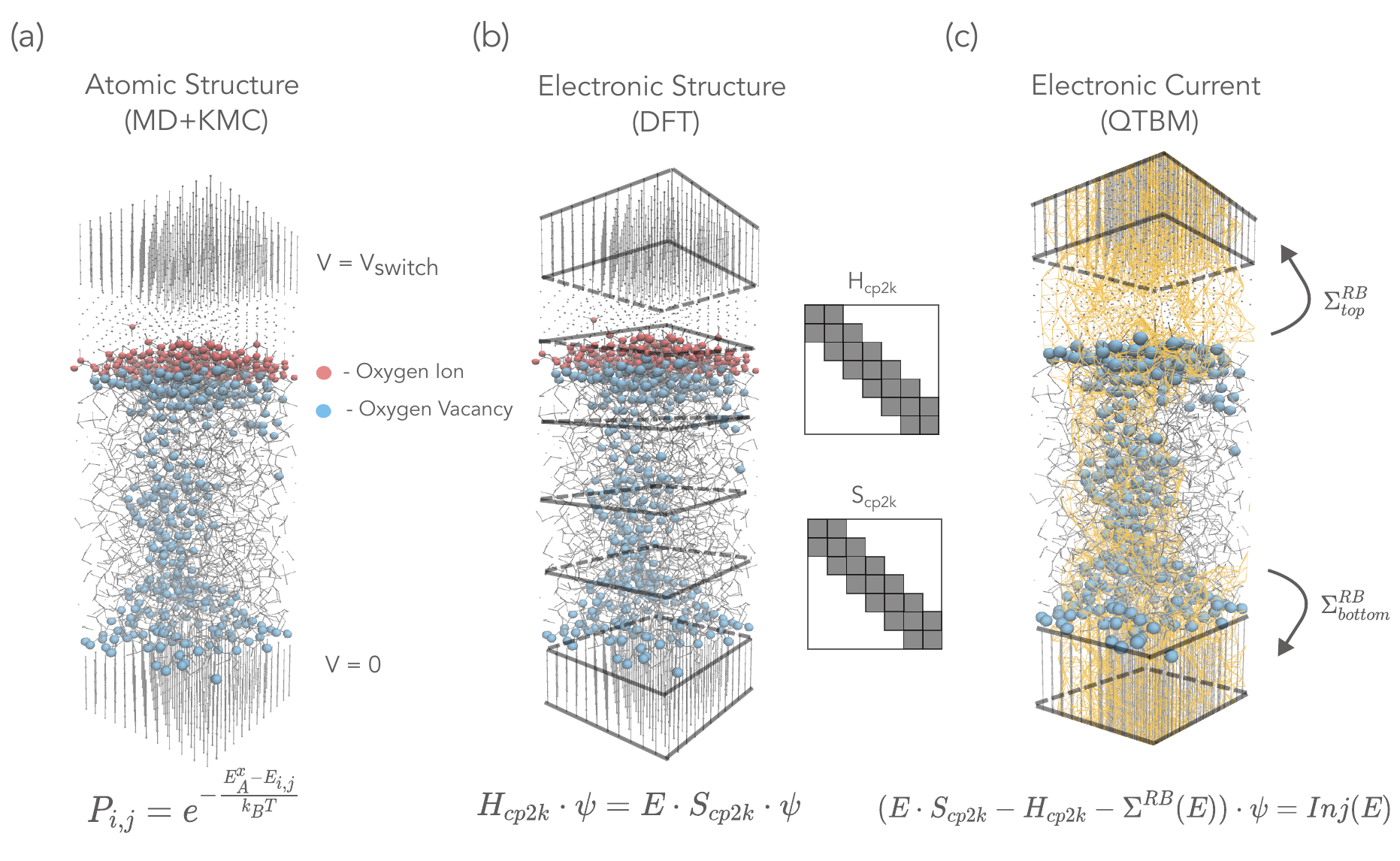}
        \end{overpic}
        \caption{Overview of the steps used to simulate the operation of VCM within our framework. Oxygen ions/vacancies are shown in artificially enlarged red/blue spheres. \textbf{(a)} Molecular Dynamics (MD) and Kinetic Monte Carlo (KMC) methods are first used to set up the structure and model the creation and movement of oxygen vacancies/ions under an applied bias of V$_{switch}$. The central equation governing this process is the calculation of event probabilities P$_{i,j}$, which depend on both the activation energy of each event type, as well as the work needed for its execution. \textbf{(b)} The final structure from the previous step is passed to the cp2k DFT code, which calculates the associated electronic structure using localized Gaussian-type orbitals. The outputs from this step are the Hamiltonian (H$_{cp2k}$) and Overlap (S$_{cp2k}$) matrices (see insets). The structure is sliced into ‘blocks’ of neighboring atoms, such as those divided here by black squares, which translate to the block tri-diagonal form of these matrices. \textbf{(c)} The electronic structure information is passed to a wavefunction-based Quantum Transport solver that computes the current flowing through the VCM under open boundary conditions, pictured here as a yellow isosurface. The black squares delimit the central device from the contacts. $Inj(E)$ here is a source term; it describes carrier injection from the contacts, which are coupled to the device through the boundary self-energies $\Sigma_{RB}$.}
\end{figure}

\subsubsection{Initial structure generation}

 The initial atomic structure of the device is pictured in \textbf{Fig.~2a}, with a 5.3 nm thick HfO$_2$ switching layer and 1.13 nm thick Ti oxygen reservoir. Its lateral dimensions are 2.69 $\times$ 2.66 nm; these dimensions were chosen to be large enough that a formed vacancy filament could be localized within the oxide without separation across periodic boundaries, and also small enough to remain computationally affordable. They are comparable to ultra-scaled experimental demonstrations of VCM \cite{Pi2018}. The thickness of the Ti layer used here is similar to the thickness of the oxidized TiO$_2$ measured via XRD in Ref. [\!\!\!\citenum{Sowinska2012}]. To generate the amorphous HfO$_2$, we apply a melt-and-quench process on monoclinic HfO$_2$ and relax the resulting atomic positions. The Ti oxygen reservoir layer and TiN contacts are attached along the transport direction.  

 Using a Nudged Elastic Band (NEB) method \cite{neb} on a subset of the device, we then calculate activation energies (E$_A$) for the following five event types: (1) oxygen vacancy/interstitial oxygen ion pair generation in bulk HfO$_2$, (2) oxygen vacancy/interstitial oxygen ion pair generation at the HfO$_2$/Ti interface, (3) oxygen vacancy diffusion, (4) oxygen ion diffusion, and (5) oxygen vacancy/ion recombination. The resulting set of E$_A$ is plotted in \textbf{Fig.~2b}. One representative atomic structure evolution for each of the processes (1)-(4) is pictured in \textbf{Fig.~2c}. During vacancy/ion pair generation, an oxygen atom leaves its lattice site and moves onto a nearby interstitial site. Vacancy diffusion occurs when an oxygen atom from the amorphous lattice moves into a vacancy location, leaving another vacancy behind, and ion diffusion occurs when an oxygen ion in an interstitial position diffuses into a neighboring interstitial position. Vacancy/ion recombination is the inverse of the generation process and typically does not present an energy barrier. Generation energies at the Ti interface are lower than those in the bulk oxide, consistent with previous studies \cite{OHara2014, Traore2018}. This behaviour can be attributed to under-coordinated interface Hf atoms being more likely to release oxygen, as well as to a different potential landscape at the interface compared to that of the bulk.  

\subsubsection{Modelling oxygen ion/vacancy motion}

To simulate the evolution of the atomic structure under an electric field induced by the switching voltage, we implemented a Kinetic Monte Carlo solver which operates directly on the atomic lattice. An overview of this model is presented as a flowchart in \textbf{Fig.~2d}. It takes the atomic structure and activation energies in \textbf{Fig.~2a-b} as inputs. At the start of the simulation, a list of `sites' is identified, which includes atomic lattice locations as well as interstitial positions facilitating oxygen ion movement. We then determine the charge states of existing ions/vacancies in the oxide, according to the vacancy charge transition model described in Ref. [\!\!~\citenum{Lee2019}]. Isolated vacancies are assigned a charge state of +2, while clustered vacancies (defined as those with at least two vacancy neighbors) are assumed neutral. Oxygen ions are in a charge state of -2, unless they neighbor a Ti atom, in which case they are neutralized. The potential at each site ($\upphi$) is calculated as the sum of a background term which accounts for the applied field and the solution of Poisson's equation over the distribution of charged species in the oxide. The background potential accounts for the atomistic connectivity of the amorphous structure and the conductivity between neighboring atomic sites, while positively/negatively charged oxygen vacancies/ions contribute additional local potentials fluctuations.

Considering this set of sites and corresponding potential, a list of events is tabulated, indexed by event type (x) and the sites involved (i, j). Events can only occur between neighboring sites; the maximum distance at which two sites are considered neighbors corresponds to the value found when calculating the activation energies of the input structure were found. For the set of E$_A$ in \textbf{Fig.~2c}, the neighboring distance is 3.5 $\text{\AA}$. Activation energies were found to increase sharply for events occurring between larger distances (Section 1 of the Supplementary Materials). For each event, a selection probability $P^x_{i,j}$ indicating the probability of event type `x' occurring between sites `i' and `j', is calculated through:

\begin{equation}
    P^x_{i,j} = \nu \cdot exp\left(-\frac{E^x_A - E_{i,j}}{k_{B}T}\right),
\end{equation}

\noindent where k$_B$ is the Boltzmann constant, T the temperature (set to 300 K), and $\nu$ the attempt frequency (set to 10THz) \cite{Zeumault2021}. In \textbf{Eqn. 1}, E$_{i,j}$ accounts for the field-induced contribution to the activation energy. As the work done to move a charge $q$ across a difference in potential $\Delta\upphi$ is  $q\Delta\upphi$, E$_{i,j}$ is computed as the sum of each moving charge multiplied by the difference in potential between its initial and final locations. The number of moving charges varies based on the event type; generation and diffusion events move only one charged species, while recombination events move two. Events are selected from the event list according to the residence-time algorithm \cite{Cox1965}. In this method, a cumulative sum of the event list $P^{1:n<N}_{i,j}$ is taken until it exceeds the total sum $P^N_{i,j}$, multiplied by a random number $0 < r_1 < 1$. The last event(n, i, j) added to the sum is then selected and assumed to occur with a duration of 

\begin{equation}
    t = -  ln(r_2) / \sum_{x, i, j} P^x_{i,j} 
\end{equation}

\noindent where $r_2$ is another random number $0 < r_2 < 1$. The event duration is therefore proportional to both the duration of the selected event, and the total duration of unselected events which could have occurred instead. 

The algorithm used here deviates from the standard KMC procedure of Ref. [\!\!\!\citenum{Andersen2019}] through an additional treatment of `simultaneous' events: in each KMC step, as defined within the dashed box in \textbf{Fig.~2d}, non-conflicting events are selected until the calculated duration of the last selected event exceeds $1/\nu$, at which point they are all executed. This allows for charged species to mutually influence each other. For example, a vacancy-ion pair generation can occur near a migrating charged vacancy, aided by its potential, without the need to `wait' for the vacancy to migrate to its lowest energy position. At low biases, the event times are typically on the order of (or above) the inverse of the attempt frequency. Hence, only one event is typically executed per step. The charge and potential are then refreshed and a new event list is compiled for the next KMC step. These KMC steps occur until the simulation timescale reaches the intended duration for which $V_{app}$ is applied. Afterwards, a device snapshot is generated.  

\begin{figure}[H]
         \centering
         \begin{overpic}[width=\textwidth, tics=10]{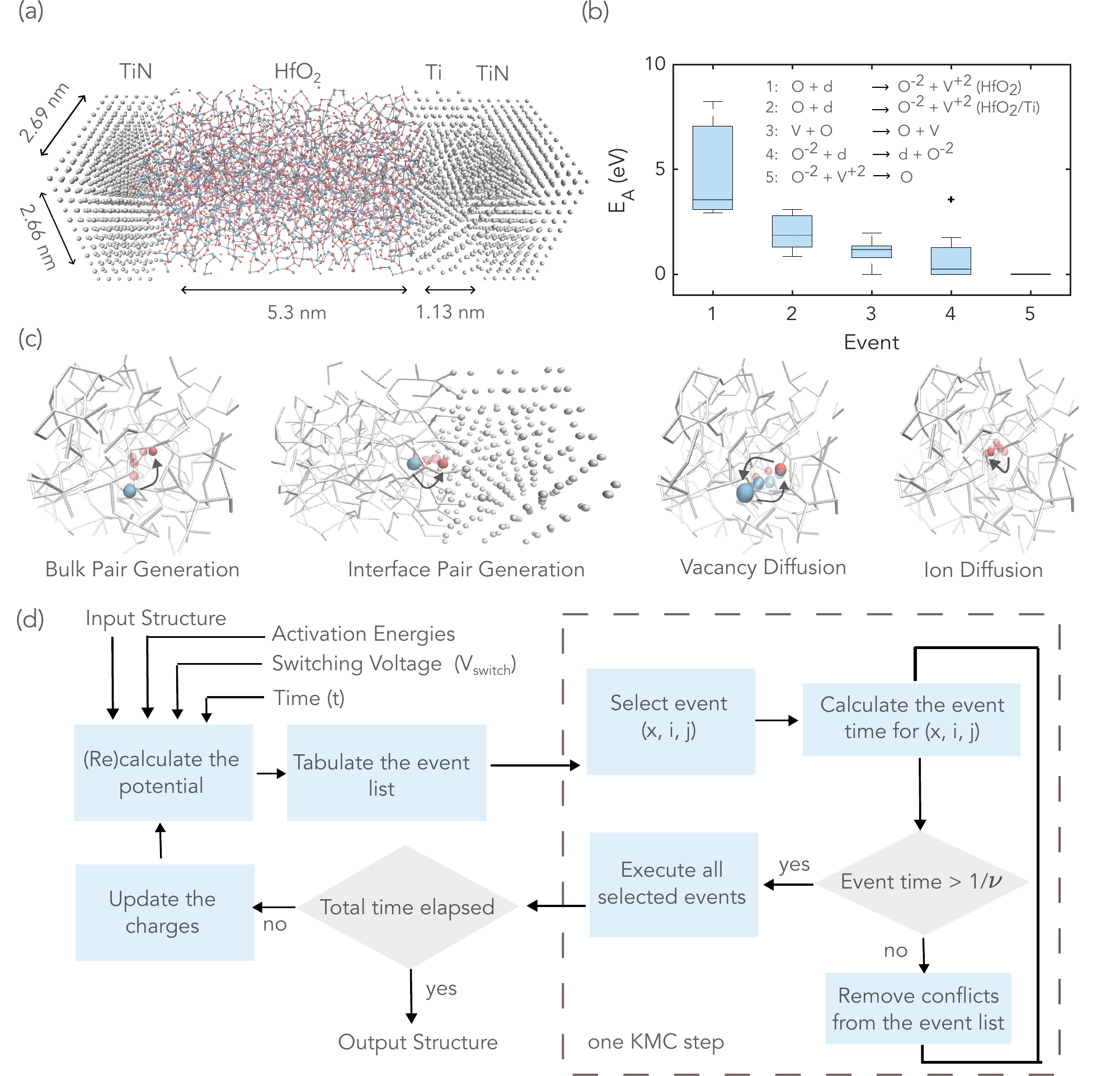}
        \end{overpic}
        \caption{Modelling the generation and movement of vacancies in a TiN/HfO$_2$/Ti/TiN device under an applied voltage of V$_{switch}$. \textbf{(a)} Atomic structure of the assembled device. \textbf{(b)} Calculated activation energy (E$_A$) ranges for the five event types detailed in the inset. `V’ denotes a vacancy while `d’ indicates an interstitial position in the lattice. \textbf{(c)} Examples of the NEB trajectories used to calculate the E$_A$ for the five event types included in the KMC model (recombination is simply the inverse of the generation event). One trajectory is shown for each event type. In each case, the ions/vacancies are pictured in red/blue. Intermediate NEB images are represented as translucent spheres while the final positions are colored solid. The HfO$_2$/Ti lattices are displayed as grey lines/spheres in the background. (d) Flowchart of the KMC method used to evolve the device structure under V$_{switch}$.}
\end{figure}

\subsubsection{Electronic current and conductance calculations}

To evaluate the conductance of the generated device snapshot, we employ the QTBM as implemented in the OMEN quantum transport solver \cite{Luisier2006}. The cp2k DFT code is first used to compute the input Hamiltonian and Overlap matrices of each snapshot, which uniquely specify the electronic structure of the device in the transport solver. Oxygen vacancies are assigned as `ghost' atoms; a basis set exists at each vacancy location, with negative onsite energies so that they cannot be occupied by electrons. This approach has been applied to defects and slab studies performed with a localized basis set and ensures a correct description of the electronic structure\cite{Saiz2018}. Oxygen ions are not included in electronic structure calculations; the states created by vacancies ultimately limit the conduction and the inclusion of ions has minimal effect on the energy-resolved transmission (see Supplementary Materials, Section 2). The electrostatic potential used for transport calculations is taken from the KMC model, by applying 0 V at the bottom contact and $V_{readout} = 0.5$ V or $V_{switch}$ at the top contact. The transport solver then computes the energy-resolved transmission function through the states available across the device. The result is integrated over to determine the final current. Finally, the device conductance ($G$) is extracted using $G = I \cdot V_{readout}$, where $I$ is the electrical current. The structures obtained by KMC simulations are not further relaxed prior to passing them to the QTBM solver. To check the validity of such an approach, we compare the transmission functions and the electrical currents before and after the structural relaxations for three distinctive oxygen vacancy concentrations in the Supplementary Materials, Section 3. We note that most of the changes in the structure occur at the interfaces, while the changes inside the bulk are minimal, resulting in small differences in current and transmission between two structures.

\subsection{Microscopic model details}

The dynamics behind the modelled filamentary growth and I-V characteristics emerge from the heuristic used for vacancies charge and the resulting potential. We provide more details on these quantities in \textbf{Fig.~3}. The charge, potential, and bond-resolved currents are calculated for three representative structures which vary in their vacancy concentration and conductance values. These structures are taken from snaphots along the structure evolution during a modelled forming process, which will be shown in the next section. 

Isolated vacancies are initially generated at the HfO$_2$/Ti interface, carry a charge of +2, and mutually repel due to Couloumb interactions. These vacancies then drift through the oxide under the applied field and begin to cluster at the bottom contact, where they lose their charge due to electrons tunneling into the defect states they create in the bandgap. Once clustered, they are no longer repulsive, allowing subsequent vacancies to drift towards them. This clustering effect continues at the center of the filament as it grows (\textbf{Fig.~3a}, middle). Once the filament has entirely bridged the oxide, the majority of the vacancies are uncharged, except for a few which remain scattered in the bulk (\textbf{Fig.~3a}, right). 

These charges, along with the atomic species information, determine the site-resolved potential pictured in pictured in \textbf{Fig.~3b} that is used to calculate E$_{i,j}$ in \textbf{Eqn. 1}. When the concentration of uncharged vacancies is low, the potential drops almost linearly across the device, with some variation stemming from the underlying atomic connectivity (\textbf{Fig.~3b}, left). Once a filament begins to form, its decreased local resistance results in it being similar in potential to the bottom contact. This creates an additional lateral potential gradient (\textbf{Fig.~3b}, middle) which attracts vacancies in the bulk to join the conductive filament, and a steeper drop in potential across the rest of the oxide which biases filament growth in that direction. Once a conductive filament has bridged the oxide, the potential within it remains constant. The applied voltage then drops mainly across the metal-oxide interfaces - to a lesser extent - at `bottleneck' sites where the filament is narrow. Note that the uncharged, clustered vacancies are immobile in the applied field. The probability of events involving their motion is therefore dependent only on the corresponding zero-field activation energy. 

This dynamic captures, on a phenomenological level, the microscopic effects of clustered defects; local fluctuations in potential caused by the appearance of charged vacancies increase the likelihood of relatively improbable events occurring in their vicinity, which results in vacancies being preferentially generated near existing charged vacancies, an effect which has been reported in previous DFT studies \cite{Gao2019, Lee2019}. The direction and orientation of filament growth is also closely dependent on the underlying atomic arrangement of the switching material. The model for the background potential recognizes conductance pathways between neighboring sites; areas of the oxide with higher density or Hf-O coordination are more likely to host vacancy filaments. This results in the existence of a few pathways along which filamentary growth is more likely to occur, the locations of which depend closely on the underlying atomic grid.

We plot the energy-resolved transmission through each of the three structures in Section 4 of the Supplementary Materials. Transmission through the bandgap of the HfO$_2$ is initially negligible, but increases as in-gap defect states are created by the presence of oxygen vacancies. \textbf{Fig.~3c-e} shows selected current isosurface values for three structures. According to the valence change mechanism, conductive pathways emerge through the orbitals of Hf atoms left under-coordinated by the presence of oxygen vacancies nearby \cite{Urquiza2021}. This effect can be seen in the insets, which show the localization of current near the vacancies. 

\begin{figure}[H]
         \centering
         \begin{overpic}[width=\textwidth, tics=10]{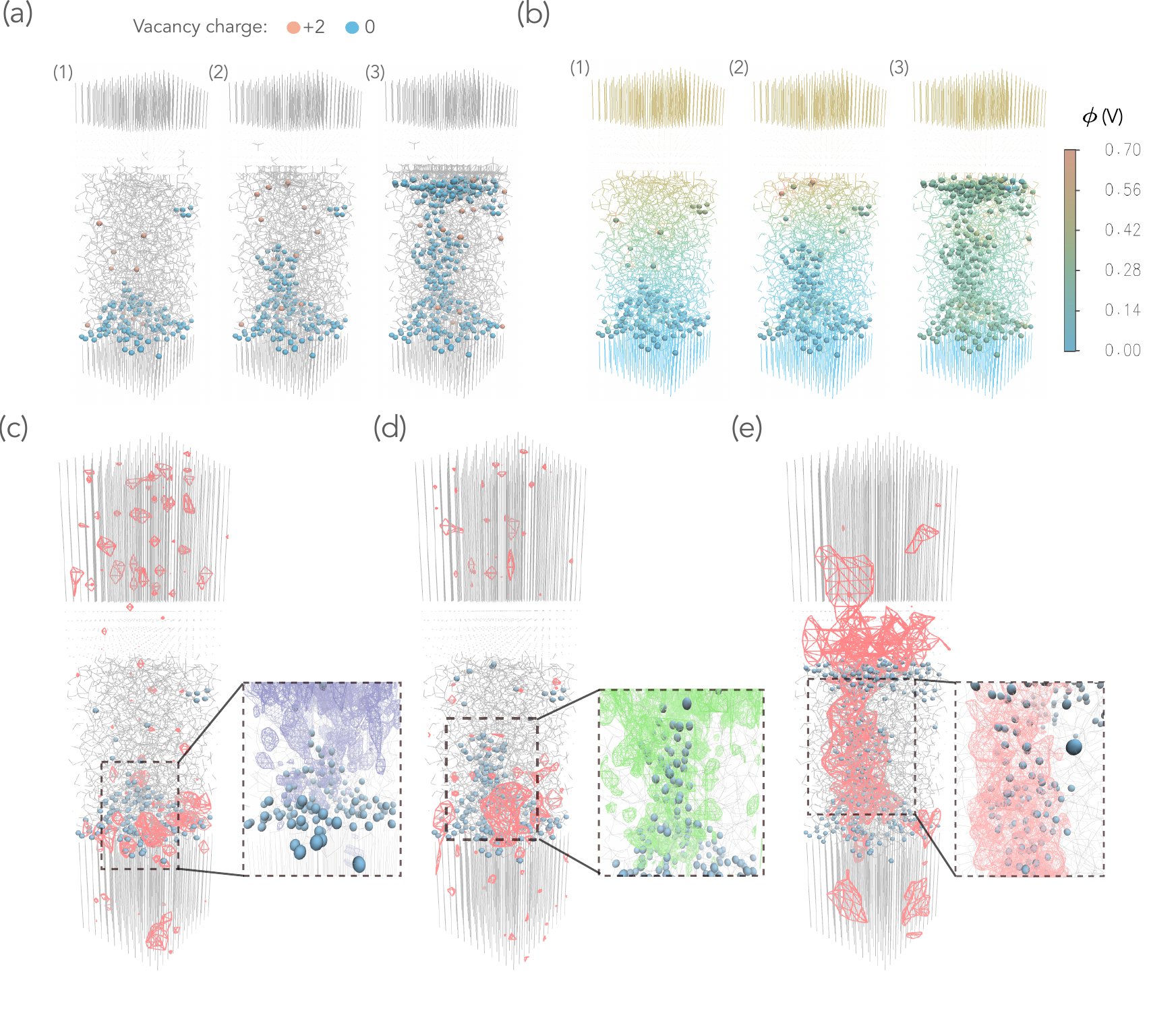}
        \end{overpic}
        \caption{Visualization of key internal quantities captured with the model, shown for three sample structures with different vacancy distributions (labelled (1), (2), (3)). Oxygen vacancy sites are pictured with spheres. \textbf{(a)} Distribution of oxygen vacancy charges. \textbf{(b)} Atomically-resolved potential, calculated for V$_{readout}$ = 0.5 V. \textbf{(c-e)} Visualizations of selected electrical current isosurfaces for the structures (1), (2), and (3). The displayed current value (dark pink isosurface) is the same in the three main plots, while the insets magnify specific areas of each structure and use different current values to highlight the electron trajectories around the oxygen vacancies, through undercoordinated Hf atoms. Note that current is conserved in all cases, but this effect cannot be read from the isosurfaces.}
\end{figure}

\subsection{Applications to Resistive Switching in HfO$_2$}

In \textbf{Fig.~4} we demonstrate the capabilities of the framework developed to model the device-level performance of the TiN/HfO$_2$/Ti/TiN VCM cell in \textbf{Fig.~2a}. We first apply an electroforming process across substoichiometric HfO$_2$. Starting from an initial, randomly distributed vacancy concentration of 5\%, we apply a forming voltage of 12 V across the two contacts and record the evolution of the atomic structure at different timescales in \textbf{Fig.~4a}. \textbf{Fig.~4b-c} quantify the vacancy concentration and device current as a function of time. The vacancy-and-ion distributions corresponding to three snapshots in the dashed box in \textbf{Fig.~4a} correspond to the structures for which the charge, potential, and current were shown in \textbf{Fig.~3}. 

Vacancies present at the beginning of the electroforming step immediately drift and pile onto the bottom contact within a few nanoseconds. This process is accompanied by a short-lived decrease in conductance as it creates a block of defect-free HfO$_2$ that acts as a tunneling barrier. Eventually, a slower generation of oxygen vacancy/ion pairs at the Ti/HfO$_2$ interface occurs at the millisecond timescale. This process is aided by oxygen ion transfer into the Ti reservoir layer. The conductance corresponding to subsequent snapshots is increased by the presence of these generated vacancies across the tunneling gap. We note that generation in the bulk rarely occurs due to this event type being energetically highly unfavorable (see \textbf{Fig.~2b}). A rapid transition into the fully formed state begins at around 10 ms thanks to a positive feedback loop created between the filament growth and the increased potential drop across the remaining insulating oxide. The forming process modelled here occurs primarily via a cascaded migration of vacancies away from the HfO$_2$-Ti interface where they are generated, consistent with experimental imaging of the dielectric breakdown process in this system \cite{Hubbard2021}. The formed structure has an oxygen vacancy concentration of 17.4\%. 

\begin{figure}[H]
         \centering
         \begin{overpic}[width=\textwidth, tics=10]{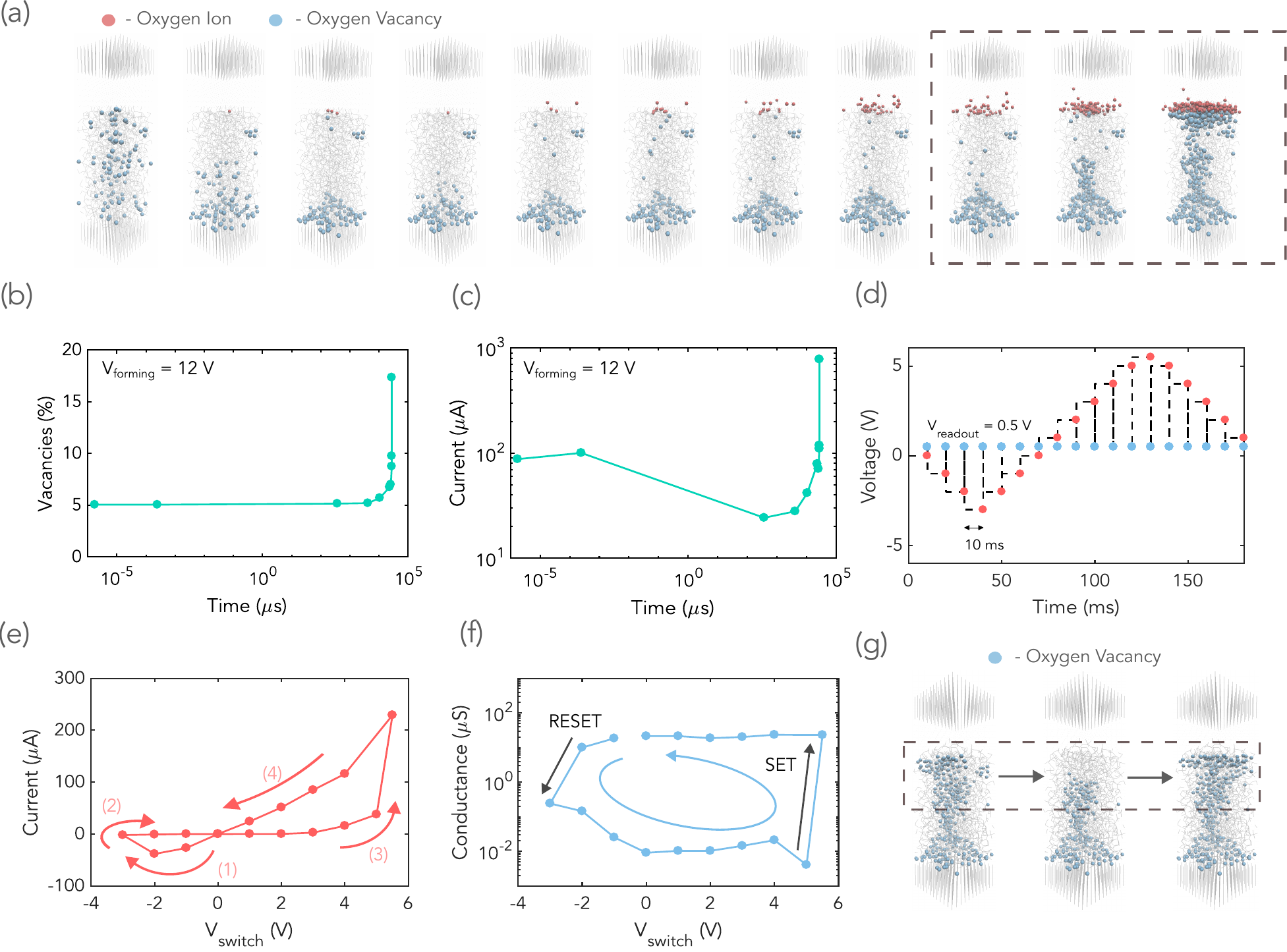}
        \end{overpic}
        \caption{Application of the simulation framework to simulate resistive switching in the device from \textbf{Fig.~2}. \textbf{(a)} Evolution of the atomic structure under a 12 V forming bias, taken at selected intermediate timepoints. The three structures in the dashed box correspond to the samples labelled (1), (2), and (3) investigated in \textbf{Fig.~3}. \textbf{(b)} Vacancy concentration/current corresponding to the timepoints at which the snapshots from \textbf{(a)} were selected. \textbf{(c)} Same as \textbf{(b)}, but for the electrical current. \textbf{(d)} Biasing scheme used to switch the device, which is initially in an LRS after forming, to an HRS and back to LRS. This scheme is applied to the data in \textbf{(e)-(f)}, where the voltages and times at which the current/conductance are readout correspond to the similarly colored points in \textbf{(d)}. \textbf{(g)} Snapshots corresponding to the initial LRS (left), HRS (middle), and final LRS (right) along the resistive switching cycle simulated in \textbf{(d)-(f)}.}
\end{figure}

After electroforming the device, we model its resistive switching in \textbf{Fig.~4d-g}. The biasing scheme used for the switching process is pictured in \textbf{Fig.~4d}, and current (conductance) is read out at the voltages and time points corresponding the red (blue) points. Since we do not explicitly set an compliance current, the voltages (V$_{RESET}$ and V$_{SET}$) required to transition the device into an HRS and LRS are identified manually through a voltage sweep. They were found to be -3 V and +5.5 V, respectively. To remove any volatility resulting from the forming process and recover stable resistive switching characteristics, we first apply 0 V to a formed device for 10 ms, and then simulate one initial switching cycle before recording the values in \textbf{Fig.~4e-f}. The I-V curve has the typical form of a memristive pinched hysteresis loop. Current in the HRS is highly non-linear as tunneling processes dominate. In contrast, the linear I-V curve in the LRS shows that a device with a bridging vacancy filament behaves as a resistor. The conductance for the LRS and HRS are stable upon removal of the applied bias (\textbf{Fig.~4f}), indicating that these states are non-volatile. \textbf{Fig.~4g} presents the vacancy distributions in the initial LRS (left), HRS (middle), and final LRS (right), showing clear modulation of a tunneling gap.

Although we show one set of results in \textbf{Fig.~4e-f}, the stochasticity of the KMC process translates directly into variations in the final conductance. To sample this variation, in \textbf{Fig.~5a}/\textbf{Fig.~5b} we simulate three RESET/SET processes starting from the same HRS/LRS structure. We then calculate the conductance of selected intermediate snapshots. The RESET processes in \textbf{Fig.~5a} occur across a wide range of timescales and access several intermediate conductance values. In general they show an initial fast drop in conductance (visible in the inset of \textbf{Fig.~5a}), followed by a slower, more gradual decrease at the millisecond timescale. Meanwhile, the SET processes in \textbf{Fig.~5b} take place entirely on the sub-millisecond timescale and are exponential in nature; the conductance increases almost instantly by over two orders of magnitude. The variations in the final HRS and LRS conductance resulting from six independent switching cycles following the biasing scheme in \textbf{Fig.~4d} are quantified in \textbf{Fig.~5c}. Non-volatile switching behavior is recovered in all cycles, but the HRS and LRS conductance vary by 89.35\% and 45.10\%, respectively. The higher variability of the HRS likely originates from the increased sensitivity of tunneling currents to the precise arrangement of vacancies at the end of the conductive filament. This effect is magnified by the small device size, which boosts the influence of fluctuations in individual atomic positions on the electrical current.

The RESET process modelled in this framework occurs via an initial `fast' mechanism at the nanosecond timescale, followed by two independent `slow' mechanisms. The fast mechanism is an instant recombination of oxygen ions into vacancy sites at the HfO$_2$/Ti interface that they can access after a single jump, resulting in the initial drop in conductance seen in the inset of \textbf{Fig.~5a}. The two `slow' mechanisms occur at the millisecond timescale. First, (1) the high field across the tunneling gap drives movement of lattice oxygen atoms at the interface into vacancy locations near the end of the filament. The vacancies thus migrate towards the interface, where they can attract and recombine with ions. In the second mechanism (2), vacancy-ion pairs generated at the interface are immediately separated by the high local electric field. The generated ions then recombine with vacancies at the end of the filament, extending the tunneling gap. The dominance of mechanism (1) in relation to mechanism (2) is dictated by the difference in activation energies between diffusion and interface generation events. Meanwhile, the SET process here is limited by vacancy-ion pair generation at the interface, followed by a migration of the generated vacancies across the tunneling gap. As the activation energy for vacancy-ion pair generation at the interface (1.66 eV) is significantly larger than the activation energy for diffusion of the generated vacancies (1.09 eV), these vacancies are immediately swept to the end of the filament, leading to the sharp increase in conductance seen in \textbf{Fig.~5b}. The switching kinetics presented here suggest that selecting oxide/reservoir bilayers engineered to have energetically favorable interface generation accelerates both the RESET and SET processes, potentially leading to a higher HRS/LRS ratio.  

We note that two key assumptions are being made in our framework. The first is related to device dimension; although the length of the oxide here is comparable to that of fabricated devices in terms of its transport direction, it is laterally constrained to a cross section of 2.69 nm x 2.66 nm. We believe that this device size is still able to capture the relevant mechanisms behind filamentary switching, as confirmed by the realistic HRS/LRS ratios produced, and represents an ultrascaled case \cite{Pi2018}. However, multi-filament growth cannot be modelled, which may be an important process in devices with gradual conductance changes. Finally, the framework accounts for field-induced switching and ignores local temperature gradients. This approximation is motivated by the assumption that the core mechanisms of bipolar switching are field-induced rather than temperature-driven \cite{Yang2012}. However, temperature is also known to play an important role in the initial forming process \cite{Padovani2015, Padovani2017} through the creation of a positive feedback loop between local heating and filament growth. It also shifts the SET and RESET processes towards lower voltages \cite{vonWitzleben2017}. This could explain why, under the set of activation energies calculated here, the magnitudes of the switching voltages ($\sim$5 V vs. $\sim$2 V) and timescales (ms vs. ns) are larger than what is encountered experimentally.

The advantages of this method reside in its ability to connect trends in measurable device characteristics directly to local changes in the atomic structure and energy landscape. Since we consider an amorphous oxide, the activation energies vary based on the local atomic environment in which they are calculated. This variability also indicates that the energy landscape can be tuned by material growth and processing conditions which alter the environment in the extended cell, for example by changing the oxide density \cite{Broglia2014}. Understanding how these conditions affect the switching kinetics thus provides insight on material- and interface-optimization. Since we directly calculate the electronic structure, the emergence of defect subbands and the interplay between hopping conduction and tunneling are also all intrinsic to the final calculated current. 

\begin{figure}[H]
         \centering
         \begin{overpic}[width=\textwidth, tics=10]{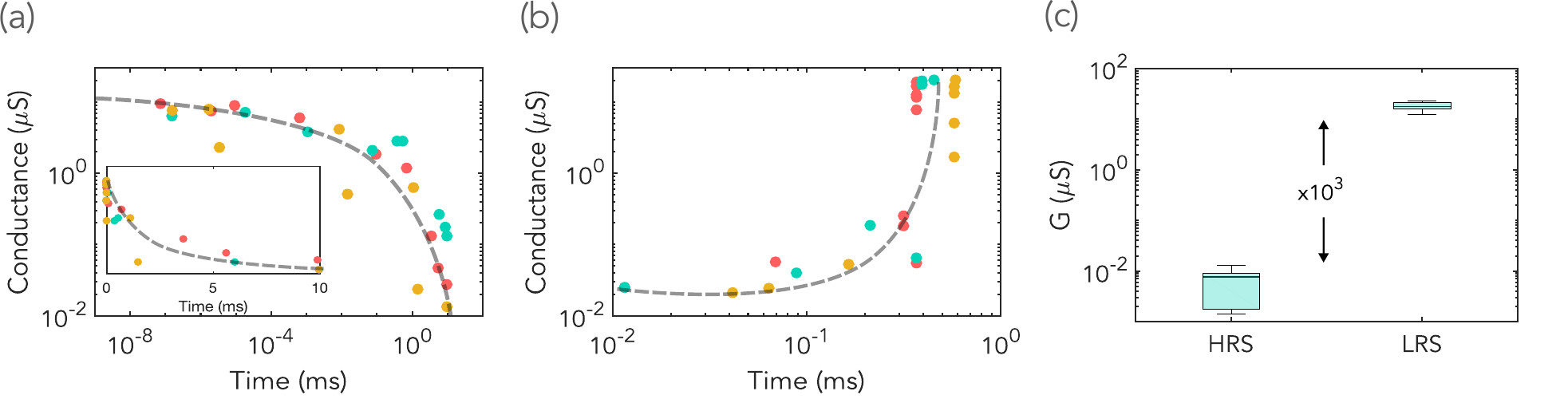}
        \end{overpic}
        \caption{Evolution of the conductance over time during the \textbf{(a)} RESET and \textbf{(b)} SET transitions labelled as such \textbf{Fig.~4f}. The insets of both plots show the same data on a linear timescale. The three colors correspond to three independent RESET/SET simulations, and the black dashed lines follow the trend as a guide to the eye \textbf{(c)} Box plot of the HRS and LRS conductances plotted for six resistive switching cycles following the biasing scheme in \textbf{Fig.~4d}, starting from the same formed device. The limits of the box indicate the 25\% and 75\% percentiles, the bars represent the minimum and maximum value, and the line in the center of each box indicates the average value.}
\end{figure}

\section{Conclusion}

We present a framework to model field-induced bipolar resistive switching in valence change memory, starting from an atomistic grid, and \textit{ab initio} energy landscape. Atomic structure perturbations are simulated using a Kinetic Monte Carlo method, which adds a stochasticity to the device performance directly related to the structure. Our framework captures non-volatile resistive switching between stable HRS and LRS conductance states stemming from the formation and rupture of a conductive filament of oxygen vacancies. Upon investigation of the dynamics of these switching processes, we find that they are driven primarily by interactions at the interface between the HfO$_2$ oxide and the Ti reservoir. The results indicate that facilitating oxygen vacancy/ion pair generation at this interface would result in an increase of the LRS/HRS conductance ratio. In conclusion, this framework provides insight on the microscopic balance of events involved in resistive switching, and offers an \textit{ab initio} methodology to compare the field-induced switching performance of different oxides and material stacks.

\section{Methods}

\subsection{Structure Setup}

To generate the amorphous HfO$_2$ oxide, we subject a block of monoclinic HfO$_2$ measuring 2.5 nm $\times$ 2.5 nm $\times$ 5.0 nm to a melt-quench-anneal process under an NVT thermostat. The structure is first heated to 3000K, then cooled to 300K at a rate of 9 K/ps, and maintained at 300K for 50 s. This initial melt-quench-anneal process is done using the LAMMPS code \cite{LAMMPS} and a ReaxFF force-field for HfO$_2$ \cite{Urquiza2021}. The purpose behind this step is to generate a realistically randomized starting structure; to remove any remaining coordination defects and unrealistic bond lengths, the structure is then fully relaxed using the cp2k DFT code and its Gaussian-type orbitals (GTO) \cite{Khne2020}. The relaxation is performed using the L-BFGS minimization method with a double zeta polarization (DZVP) basis to accurately capture bond lengths. The plane-wave cutoff is set to 500 Ry, while a cutoff of 60 Ry is used for mapping the GTOs onto the grid. Convergence criteria of $4.5\cdot 10^{-4}$ Ha/Bohr for forces and $3\cdot 10^{-3}$ Bohr for the geometry change are used. The reciprocal space is sampled at the $\Gamma$-point only, due to sufficient size of the structure. The final density of the relaxed oxide is 9.22 g/cm$^3$, which falls in the range of experimentally reported densities (5-12 g/cm$^3$) \cite{Martinez_2007}.

 The amorphous oxide is then attached to the left TiN contact (the inactive electrode), which is strained by 6.1 \% and 5.0 \% in the y- and and the z-direction, respectively, to match the cell size of the oxide. The spacing between the TiN and oxide interfaces is optimized such that it results in the minimum self-consistent field energy in cp2k. This method captures both the relaxed atomic positions and interface spacing while avoiding the computational cost associated with a relaxation of the entire device. Next, around 1.13 nm of hexagonal Ti is attached to the right side (strained by 1.3 \% and 5.0 \%), followed by 1.05 nm of TiN, with the HfO$_2$/Ti and Ti/TiN interface spacings similarly optimized. Details of the inter-layer spacing optimizations are given in Section 5 of the Supplementary Materials. To evaluate an effect of strain in the Ti layer on the transmission function through the device, we compare the DOS of the strained and the relaxed Ti in Supplementary Materials, Section 6. The qualitative difference between two DOS plots is minor, leading us to conclude that the added strain has a minimal effect on the results. 
 
To complete the structural inputs to the KMC model, a set of interstitial positions which oxygen ions could occupy following a generation event has to be identified. While the identification of interstitial positions in a crystalline structure is intuitive, it is challenging in the case of an amorphous structure. In our model, the interstitial positions are defined as those where the coordination number of an O ion is altered due to the breakage of one or more Hf-O bonds and/or due to the creation of an O-O bond. The identification process proceeds as follows: initially a grid of 1.2 $\text{\AA}$ inter-spacing distance and of same dimensions are the simulation box is created. Subsequently, a set of interstitial positions is defined as a grid subset that satisfies the following set of conditions: (1) a minimal distance from any Hf atom of 2.2 $\text{\AA}$, (2) a minimal distance from any O atom of 1.2 $\text{\AA}$, and (3) a minimal distance from any Ti atom of 1 $\text{\AA}$. Finally, the interstitial positions are relaxed by keeping lattice atoms fixed and by using the same set of calculation parameters, except for the usage of short-range single-zeta valence (SZV) basis set due to increased calculation complexity, increased by the presence of interstitial atoms. Such a procedure allows for a simultaneous identification of interstitial sites in bulk HfO$_2$ as well as in the oxygen reservoir layer at the HfO$_2$-Ti interface and in the Ti layer. The resulting set of positions is shown in Section 7 of the Supplementary Materials. We note that due to the energetic cost of generating a vacancy/ion pair in bulk HfO$_2$, the majority of these interstitial positions remain unoccupied during the simulation.

\subsection{Calculation of the Activation Energies and Amorphous Oxide Structure}

Since the structure is amorphous, the activation energy for events varies based on the local environment. For each event, we consider five different paths (three in the case of vacancy generation at the HfO$_2$-Ti interface) of such an event occurring at different places along the oxide. These energies are averaged for the subsequent KMC simulations. The energies are calculated using the NEB method as implemented in the cp2k code. Due to high computational costs of NEB calculations, a smaller structure of HfO$_2$ is generated (1.65 nm $\times$ 1.61 nm $\times$ 1.65 nm), following the procedure for amorphous structure generation and relaxation described in the `Structure Setup' section. To calculate the activation energies for vacancy generation at the active electrode, a structure with a HfO$_2$-Ti interface is created and optimized. We note that the Ti in this structure adopts a cubic fcc phase, rather than a hexagonal hpc phase used in the device simulations. Such a transition has been experimentally observed \cite{Yang2018} and is not expected to have an influence on the model. We indeed consider the vacancy-ion pair generation in HfO$_2$ in the close vicinity of a Ti layer, but not inside the Ti layer where the inter-atomic arrangement of Ti atoms would matter. For each NEB calculation, 7 images are considered to identify the saddle point, including the initial and the final structure, that are relaxed prior to performing the NEB calculation. To prevent the vacancy-ion recombination from happening, the positions of the interstitial O atom and the close Hf atoms are kept fixed during the relaxation of the structures used as the end points in the NEB calculations for the vacancy generation and ion diffusion events. Energy profiles of all NEB calculations performed in this work are given in Section 8 of the Supplementary Materials. 

\subsection{Calculation of the Potential in the KMC model}

The site-resolved potential can be decomposed into two contributions; one from the external field (`background'), and one from the potential created by charged species in the oxide. We thus solve for the potential $\upphi$ for every position `i', where:

\begin{equation}
    \upphi_i = \upphi_{background, i} + \upphi_{charge, i} 
    \label{eqn:potential}
\end{equation}

To solve $\upphi_{background}$, we define an adjacency matrix $K$ with non-zero entries for all site-neighbor connections (i, j). Uncharged (clustered) vacancies are considered `conductive', so that the entries of K can be approximated as:

\begin{equation}
    K_{i,j} = 
    \begin{cases}
        1, & \text{if } i, j\in \text{conductive or } i, j\in \text{contacts}\\
        1\times 10^{-8}, & \text{otherwise}
    \end{cases}
\end{equation}

 We then solve \textbf{Eqn. 5} for $\upphi_{oxide}$, which gives us the site-resolved potential within the oxide. 

\begin{equation}
    \begin{bmatrix}
   K_{bottom, contact} & K_{bottom, interface} & 0 \\
   K'_{bottom, interface} & K_{oxide} & K_{top, interface} \\
   0 & K'_{top, interface} & K_{top, contact}
   \end{bmatrix} \cdot \begin{bmatrix}
   \upphi_{bottom, contact} \\
   \upphi_{oxide} \\
   \upphi_{top, contact}
   \end{bmatrix} = \begin{bmatrix}
   I_{bottom, contact} \\
   0 \\
   I_{top, contact}
   \end{bmatrix}
\end{equation}

\begin{equation}
    \upphi_{oxide} = [K_{oxide}]^{-1}\cdot (K'_{bottom, interface}\cdot\upphi_{bottom, contact} + K_{top, interface}\cdot\upphi_{top, contact})
\end{equation}

\noindent Here, $K_{bottom, contact}$ and  $K_{top, contact}$ are the submatrices of K corresponding to connections between sites in the contacts. $K_{bottom, interface}$ and $K_{top, interface}$ correspond to connections between sites assigned as `contacts' and sites within the central oxide. $I_{bottom, contact}$ and $I_{top, contact}$ represent the current flowing into the bottom and top contacts, respectively. These quantities are not explicitly calculated, since the solution of $\upphi_{oxide}$ is independent of them. It relies only on the boundary potentials ($\upphi_{bottom, contact}$ and $\upphi_{top, contact}$) and the net current flowing into sites within the oxide (which is 0 due to current conservation). The boundary potentials are the external potential applied to every site in the contacts. This corresponds to `0' for the bottom contact sites, and `V$_{switch}$' for the top contact sites. The entries in $\upphi_{oxide}$ thus fall within the range [0, V$_{switch}$]. From this, $\upphi_{oxide}$ in \textbf{Eqn. 6} gives us the site-resolved potential within the oxide. Finally, the assembled background potential is:

\begin{equation}
    \upphi_{background} = [\upphi_{bottom, contact}, \upphi_{oxide}, \upphi_{top, contact}]
\end{equation}

The second term in \textbf{Eqn. \ref{eqn:potential}}, $\upphi_{charge, i}$ accounts for the existence of other charged species in the oxide, namely the charged vacancies and ions. For each neutral site `i' we calculate a short-range solution of Poisson's equation in three dimensions\cite{Lee2009}:

\begin{equation}
    \upphi_{i} = \sum_{c} erfc\left(\frac{r_{i,c}}{\sigma\sqrt2}\right)\frac{kq_{c}}{r_{i,c}},
    \label{eqn:psi}
\end{equation}

\noindent where `c' indexes all charged species, k is the Coulomb constant, and $\sigma$ is set equal to the neighbor radius used thoughout the KMC model (3.5 $\text{\AA}$). Then, for each charged site `c', we find the potential as:

\begin{equation}
    \upphi_{c} = max(\upphi_{n.n.} - \upphi_{c, n.n.}),
\end{equation}

\noindent where $\upphi_{n.n.}$ runs over the neighbors of site `j', and $\upphi_{c, n.n.}$ is the potential on the neighboring site resulting from the charge on site `c', which contributes to the sum in \textbf{Eqn. \ref{eqn:psi}}. 

To diminish self-interaction errors in the potential, we implement the `exclusion' scheme proposed in Ref. [\!\!~\citenum{Li2017}] as a correction term when calculating $E_{i,j}$ in \textbf{Eqn. 1}. Namely, we compute the difference in potential between two sites `i' and `j' as $\upphi_{i} - \upphi_{j} + \upphi_{i, j}$, where $\upphi_{i}$ and $\upphi_{j}$ are the potentials at sites `i' and `j', and $\upphi_{i, j}$ is the potential at site `j' resulting from the charge at site `i'. The latter term is zero in the case of two neutral sites. This scheme ensures that in the case of zero bias, $E_{i,j}$ approaches zero and $P^x_{i,j}$ depends only on the zero-field activation energy of the corresponding event. 

This solution is then added to the background potential at each site. We note that the potential arising from the solution of Poisson's equation is independent of the background potential or the applied field. The total potential calculated at either V$_{switch}$ or V$_{readout}$ is used later to calculate the current or conductance.

\subsection{Computation of the Hamiltonian and Overlap matrices}

We use the cp2k code to compute the Hamiltonian (H$_{cp2k}$) and Overlap (S$_{cp2k}$) matrices of each snapshot, which solve the Schroedinger's Equation with a non-orthogonal basis set: $[H_{cp2k}]\Psi = E[S_{cp2k}]\Psi$. To generate these matrices, the orbitals for all the atoms in the device are described by a short-range single-zeta valence (SZV) basis set, which was found sufficient to capture trends in the final transmission (See Ref. [\!\!\citenum{Ducry2020}]). 

Using a higher number of basis functions per orbital would lead to more accurate results for a larger transport energy window, but is computationally unfeasible for the system sizes considered here. We also use the Orbital Transform (OT) \cite{orb_trans} minimization scheme, and converge the self-consistent field until the gradient of the energy with respect to any of the molecular orbitals is under 1$\times$10$^{-6}$ Ry. Periodic boundary conditions are applied in all directions, but we note that the filament in each of the snapshots is fairly localized in lateral directions, such that the current is unlikely to flow across multiple unit cells.

To correct for the bandgap underestimation from DFT, we employ a DFT+U approach \cite{dft+u} with parameters $U_{Hf} = 7\ eV$ and $U_{O} = 10 \ eV$ applied to the 5d orbital of Hf and the 2p orbital of O, respectively. These parameters are identified by matching the calculated bandgap to the experimental value for amorphous HfO$_2$, that is reported to be in the 5.3 - 5.7 eV range \cite{Perevalov2007, Kaiser2022}. The evolution of the band gap as a function of U parameters is given in Section 9 of the Supplementary Materials.

\subsection{Quantum Transport}

Coherent transport occurs through states $\psi(E)$, which are found by solving: \begin{equation}
    (E\cdot S_{cp2k}-H_{cp2k}-\Sigma^{RB}(E))\cdot\psi(E)=Inj(E)
\end{equation}

\noindent Here, $Inj(E)$ describes carrier injection from the contacts, which are coupled to the device through $\Sigma^{RB}(E)$ \cite{Luisier2006}. The Landauer-Buttiker equation is then used to calculate the energy-resolved current. We note that this method models coherent transport; the effects of phonon scattering are not included as they incur a high computational cost. Due to the homogeneity of the electronic structure in an amorphous oxide, an inclusion of scattering is unlikely to significantly change the carrier velocity. However, it may increase the number of carriers able to flow through the oxide by providing a mechanism for transport between defect states at different energies. In the case of trap-assisted tunneling through a defective oxide, we expect this effect to be most noticeable at lower defect densities, and saturate in the LRS where conduction happens primarily via transport through defect sub-bands.

\begin{acknowledgement}

All authors acknowledge funding from SNSF Sinergia (grant no.
198612), and computational resources
from the Swiss National Supercomputing Center (CSCS)
under project 1119. M.L. and M.M. acknowledge funding from the Werner Siemens Stiftung Center for Single Atom Electronics and Photonics. M.K. acknowledges the Natural Sciences and
Engineering Research Council of Canada (NSERC) Postgraduate Scholarship (PGS-D3).

\end{acknowledgement}

\begin{suppinfo}

The Supporting Information is available at [link], and includes details behind the NEB activation energy, calculations, details behind device structure setup, DFT+U correction parameters, plots showing the effect of oxygen ions on the transmission, and data for smaller structures after structural relaxation of vacancy profiles.

\end{suppinfo}

\newpage 
\bibliography{manuscript}

\newpage

\end{document}